**Маркова Оксана Миколаївна**
старший викладач кафедри комп'ютерних систем та мереж
ДВНЗ «Криворізький національний університет», м. Кривий Ріг, Україна
*kissa_oks@mail.ru*

**Семеріков Сергій Олексійович**
професор, доктор педагогічних наук,
завідувач кафедри фундаментальних і соціально-гуманітарних дисциплін
ДВНЗ «Криворізький національний університет», м. Кривий Ріг, Україна
*semerikov@gmail.com*

**Стрюк Андрій Миколайович**
кандидат педагогічних наук, докторант
Інститут інформаційних технологій і засобів навчання НАПН України, м. Київ, Україна
*andrey.n.stryuk@gmail.com*

# ХМАРНІ ТЕХНОЛОГІЇ НАВЧАННЯ: ВИТОКИ

**Анотація.** Метою дослідження є розгляд еволюції концепції комп'ютерної послуги у роботах зарубіжних дослідників 1959–66 рр. Уперше введено у вітчизняний науковий обіг результати А. О. Манна й розширено огляд результатів Д. Ф. Паркхілла щодо концепції комп'ютерної (інформаційної) послуги. Доведено функціональну ідентичність концепції комп'ютерної і хмарної послуги, уточнено першоджерела моделей надання хмарних послуг. Наведено трактування поняття «хмарні технології навчання». Зроблено висновок про неперервність розвитку хмарних технологій за останні 55 років і їх зв'язок із розвитком ІКТ у цілому. Результати дослідження надають можливість визначити перспективи розвитку хмарних технологій у цілому і хмарних технологій навчання зокрема.

**Ключові слова:** комп'ютерна послуга; комунальні обчислення; хмарні технології; хмарні технології навчання.

## 1. ВСТУП

**Постановка проблеми.** Теоретичною основою хмарних технологій є концепція «комунальних обчислень» (utility computing), сутність якої у 1961 р. Дж. Маккарті (John McCarthy, 1927–2011) виклав у доповіді, присвяченій сторіччю Массачусетського технологічного інституту [1], розглянувши комп'ютерні ресурси (обчислювальні, зберігальні та інші) як вимірювані і гнучко дозовані послуги на зразок тих, що надають оператори зв'язку: «Комп'ютерні ресурси можуть бути організовані як комунальні послуги на зразок телефонної системи. ... Кожному абоненту такої послуги необхідно сплачувати лише за спожите, проте він буде мати доступ до усіх мов програмування на великій кількості систем. ... Деякі абоненти також можуть надавати послуги іншим. ... Комунальні обчислення мають стати основою нової та важливої індустрії» [4, с. 2].

Продаж комп'ютерних послуг у 1960-х рр. не був звичайною бізнес-моделлю: як зазначає Т. Хей (Thomas Haigh) [8, с. 8], у деяких випадках контракти на встановлення комп'ютерного обладнання передбачали надання програмних послуг і встановлення стандартних пакетів. Із зростанням інтересу до мережного і віддаленого доступу до комп'ютерів очікувалось, що така бізнес-модель стане звичною. Основою побудови «комунальних комп'ютерів» (computer utilities) вважались системи розподілу часу (коли декілька користувачів отримували доступ у режимі реального часу до одного комп'ютера). За такої моделі тисячі користувачів могли бути абонентами гігантських мереж, використовуючи термінали для доступу до комп'ютерного обладнання і програмного забезпечення, що виконувалось на віддалених комп'ютерах.

Із часом трактування «computer utility» як «комп'ютерної послуги» змінилось з власне послуги, що надається, на її програмне забезпечення. Проте початкове трактування в останнє десятиріччя знову набуває актуальності, тому доцільним є дослідження генезису й еволюції хмарних технологій як різновиду комп'ютерної послуги.

**Аналіз останніх досліджень і публікацій.** У щорічному звіті компанії Gartner за 2009 рік [2] хмарні технології були названі черговим супер-концептом серед ІКТ, на який покладено надзвичайні сподівання («Cloud Computing is the latest super-hyped concept in IT»). Згідно звіту, хмарні технології хоча і вважаються дуже простою ідеєю – отримання послуг з «хмари», та є багато питань, що стосуються видів хмарних технологій чи масштабів їх розгортання, які роблять їх не такими простими. О. В. Чорна, використовуючи фірмову технологію Gartner – цикл надочікувань – для виявлення тенденцій розвитку хмарних технологій, доходить висновку про необхідність діяти «на випередження»: системного і глибокого «вивчення ... іноземного досвіду у цій сфері надає змогу усунути ймовірні недоліки до їх практичного виявлення і таким чином уникнути небажаних наслідків при їх ефективному використанні» [22, с. 6].

Широка розрекламованість хмарних ІКТ призвела до появи великої кількості однотипних публікацій, що спрощено висвітлюють історію їх появи і розвитку в дусі відповідного історичного розділу популярної статті у Вікіпедії [21], який з'явився у березні 2012 року. Намагаючись виправити цю ситуацію, автори колективної монографії [20] першу частину II розділу присвятили історії виникнення хмарних послуг, проте наукові результати, що стосуються витоків хмарних технологій, подані лише оглядово. Тому **метою статті** є розгляд еволюції концепції комп'ютерної послуги у роботах зарубіжних дослідників 1959–66 рр.

## 2. РЕЗУЛЬТАТИ ДОСЛІДЖЕННЯ

У 1959 році А. О. Манн (Alan O. Mann, 1907?–після 1974) у статті [12], прогнозуючи розвиток менеджменту систем управління (нині відомого як «контролінг»), зокрема, якості послуг, вказував на 4 головні фактори, що спостерігались у розвитку електричних, телефонних, телеграфних та інших систем бізнесу з продажу (постачання послуг):

1) розмір інвестицій, необхідних для кожної окремої системи;
2) потенційне перекриття і дублювання елементів системи із збільшенням кількості окремих систем;
3) невід'ємна для електронних систем властивість зменшення вартості опрацювання одиниці із зростанням швидкості, обсягу та потужності;
4) зростання вартості утилізації кожної окремої системи.

Такі системи А. О. Манн називає системами комунальних послуг (public utility), особливостями яких є:

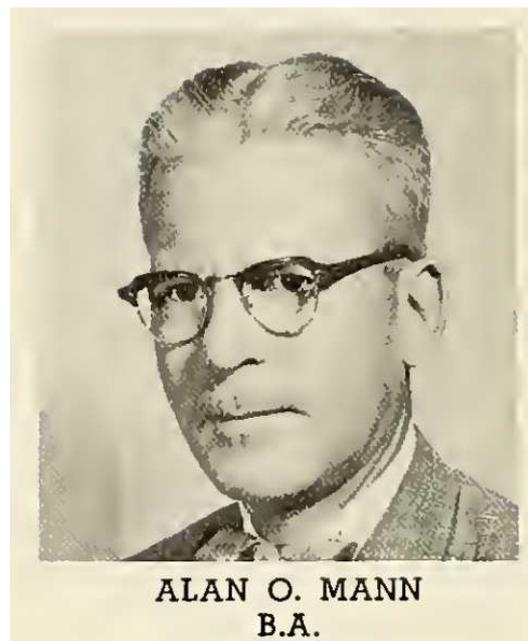

*Алан О. Манн як викладач Коледжу Ла Саля (La Salle College) у Філадельфії (з щорічника за 1960 рік)*

- наголос на максимізацію обсягу послуг за мінімізації прибутку за безперервного й ефективного ведення бізнесу;
- зацікавленість громадськості у постійному детальному нагляді, висока ступінь регулювання відповідності бізнесу суспільним інтересам;
- обслуговування всіх бажаючих за розумною ціною і без дискримінації;
- централізоване постачання послуги;
- наголос на задоволенні потреб користувачів послуги;
- акцент на ініціюванні радикальних бізнес-інновацій;
- виховання відданості працівників службі;
- вплив на послугу зв'язків із громадськістю;
- акцент на наукові дослідження і технологічне лідерство;
- послуга має надаватися за вимогою і приносити прибуток після капітальних вкладень.

До переваг використання комп'ютерних систем як комунальних послуг А. О. Манн відносить такі:

а) можливість використання розробленого програмного забезпечення на різних комп'ютерних системах, у тому числі – неіснуючих на час розробки програми;

б) зростання кількості комп'ютерних систем покращує надання послуги;

в) вартість послуги зменшується із зростанням розміру і потужності системи;

г) резервна потужність комп'ютерних систем, необхідна для задоволення пікових потреб користувачів, нарощується швидше й дешевше за інші типи систем за рахунок їх модульності;

д) диверсифікація послуги між постачальниками відбувається прозоро для її споживача;

е) додавання нових компонентів комп'ютерної системи збільшує її потужність за меншу вартість порівняно з попередніми компонентами;

ж) об'єднання наукових обчислень й опрацювання бізнес-даних у спільній системі загального призначення сприяє економії коштів так само, як і надання різним споживачам стандартизованих послуг.

Попри це, є й деякі історичні передумови для розгляду комп'ютерних послуг як комунальних.

1. Багато з існуючих комунальних послуг розвивались за такою ж схемою, як і комп'ютерний бізнес:

а) зростаюча кількість користувачів організується у спільні підприємства і кооперативи;

б) починається продаж надлишкових потужностей і послуг;

в) зростаючі потреби користувачів не можуть бути задоволені швидко й адекватно за місцем вимоги, що приводить до необхідності розвитку постачальника послуги.

У комп'ютерному бізнесі є вимоги, аналогічні вимогам користувачів комунальних послуг: більш високі швидкості, більші обсяги збережуваних даних, краще введення-виведення, підтримка мов загального призначення і спрощеного програмування, менші витрати на заміну комп'ютерного обладнання і його перепрограмування, канали передавання даних більш низької вартості з радикально більшою надійністю і швидкістю, надання високоякісних послуг, а також зниження цін у цілому.

2. Існує безсумнівна зростаюча тенденція до об'єднання комп'ютерних систем з мережами зв'язку, адже значна частина всіх опрацьовуваних даних поступає через канали їх передавання. Отже, усе більше й більше традиційних постачальників зв'язку надають лінії передавання даних як елементи інтегрованої системи опрацювання даних. Безперечно, ця тенденція буде зростати з боку як різноманітності й масштабу, так і з боку спільної вартості послуг передавання даних, адже зв'язок нині вже є комунальною послугою.

3. Розповсюдженою практикою в комп'ютерному бізнесі стає оренда, а не продаж обладнання, із супутнім наданням обслуговування, запасних частин, навчання, резервного копіювання, а також численних послуг, що охоплювала орендна платня. Це є ще однією важливою історичною характеристикою комунального бізнесу.

У 1959 році було важко уявити будь-яку перспективу для окремої людини як користувача великомасштабної комп'ютерної системи, проте А. О. Манн наголошує, що «такі щоденні вимоги, як обробка даних або інформаційний пошук потребують оренди машинного часу» [12, с. 257], тому найперспективнішою комп'ютерною послугою вважає доступ до комунальних мереж передавання й опрацювання даних, об'єднання яких нині відоме як Інтернет.

Протилежністю «комунальним обчисленням» на основі великих, потужних, швидкісних центрів опрацювання даних, які обслуговують велику кількість клієнтів, А. О. Манн вважає засоби «персональних обчислень» – сучасні персональні комп'ютери, вказуючи, що зростання їхніх характеристик може призвести до відсутності необхідності у «комунальних обчисленнях», проте сам же й наводить контртезу – зростаюча складність зв'язків і комунікацій змушує об'єднуватись для ефективного опрацювання даних: «Цілком можливо, що розвиток комп'ютерних систем буде подібний практиці, яка існує в електричній та телефонній системах, коли деякі фірми або організації-споживачі закуповують та експлуатують свої власні системи, але в той же час користуються послугами інженерних мереж. ... Такий розвиток подій може бути гарним для користувачів» [12, с. 258].

До технологічних проблем «комунальних обчислень» – надання комп'ютерних послуг – А. О. Манн відносить:
– нерозробленість методів високошвидкісного високонадійного зв'язку;
– відсутність стандартизації устаткування, що має працювати тривалий час;
– необхідність стандартизації програм і даних засобами автоматизації програмування мовами загального призначення.

Реалізація запропонованої концепції мала надати такі переваги:
– для організацій: можливість споживання більшої кількості комп'ютерних послуг меншої вартості;
– для постачальників комп'ютерного обладнання: перспективи для збільшення швидкості, ємності, потужності та переходу від друкованих документів до пристроїв зчитування символів і візуальних дисплеїв;
– для розробників програмних систем: засоби уніфікації використовуваних мов програмування, трансляторів тощо;
– для користувачів: рівний доступ до програм і даних, розв'язання проблем немобільності програм і обладнання, застосування сучасних технологій опрацювання даних та інші засоби задоволення їх зростаючих потреб.

Суттєвий внесок у дослідження проблеми комп'ютерної послуги вніс канадський ІТ-фахівець Д. Ф. Паркхілл (Douglas F. Parkhill, 1923?–1995?), огляд однойменної книги якого, виданої 1966 року, виконав американський математик, автор відомої «ділеми ув'язненого» та піонер дослідження операцій і системного аналізу М. М. Флуд (Merrill M. Flood, 1908–1991) [3]: «Ця книга є особливо важливою для тих, хто має досвід роботи з комп'ютерними системами розподілу часу, як корисний приклад інформаційних послуг. Автор починає з короткого, але значущого аналізу історії обчислювальної техніки. ... Його історичний аналіз впливу ... ранніх «мульти-» систем допомагає визначити перспективу для більш нових розробок «комп'ютерних послуг». ...

Для пояснення основних характеристик існуючих систем розподілу часу і спільного використання файлів у комп'ютерних системах детально обговорюються дев'ять «ранніх комп'ютерних послуг». ... Автор визначає Computer Sciences Corporation (REMOTRAN) як первісток «публічної комп'ютерної послуги загального призначення», зазначаючи, що «історики можуть коли-небудь розглядати 14 грудня 1964 року, як день, коли комп'ютерна послуга, нарешті, досягла повноліття» ...

Автор наводить власний аналіз імовірного характеру майбутніх комп'ютерних послуг через екстраполяцію існуючих комп'ютерних технологій (апаратного і програмного забезпечення), а також основних економічних, соціальних, технічних і правових проблем, які необхідно розв'язати для швидкого просування комп'ютерних послуг. ...

Автор книги ... вказує великі переваги MAC-подібних систем [Multiple Access Computer – комп'ютер спільного доступу] для введення даних, редагування збережених

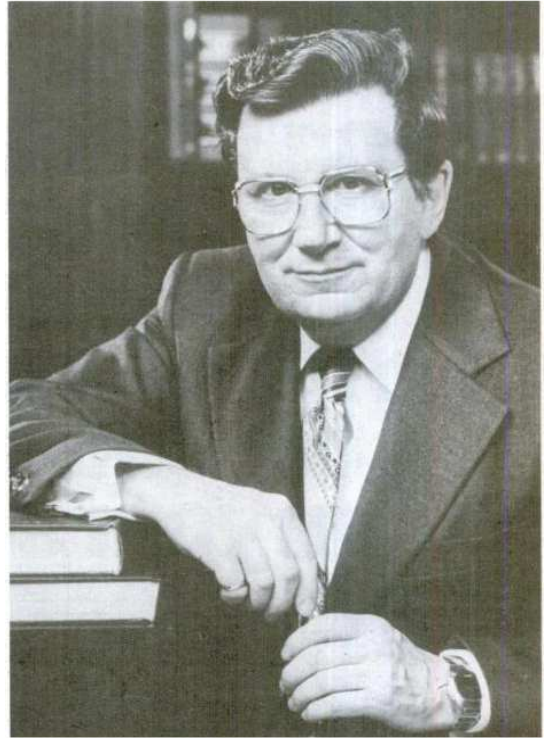

*Дуглас Ф. Паркхілл під час роботи над Telidon (фотографія за 1981 рік з газети Computerwold)*

файлів та легкого обміну файлами програм і даних між користувачами».

М. М. Флуд не поділяє занепокоєння Д. Ф. Паркхілла щодо можливих негативних наслідків розвитку комп'ютерних послуг: загрози конфіденційності, можливе технологічне безробіття, можливості для політичного контролю і багато інших потенційних небезпек можуть і не здійснитися, якщо будуть укладені угоди з усіма зацікавленими сторонами: державою, бізнесом, абонентами послуг. Сам Д. Ф. Паркхілл власне бачення комп'ютерної послуги реалізував, починаючи з 1969 року, у Міністерстві зв'язку Канади під час розробки і впровадження відеотекс-телетекстової системи Telidon, основна частина якої – комунікаційний протокол передавання даних у телевізійному сигналі (NABTS) – використовувалась до 2014 року в MSN TV.

Серед ранніх дослідників з Массачуссетського технологічного інституту, що працювали над згадуваним

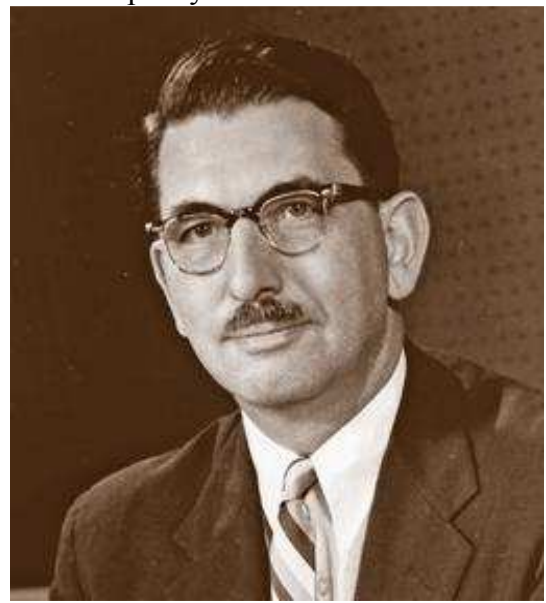

*Мерріл М. Флуд у 1956-1976 рр. – професор Мічиганського університету*

Д. Ф. Паркхіллом Project MAC, виділимо М. Грінбергера (Martin Greenberger, нар. 1931), який був піонером проектування систем розподілу комп'ютерного часу і даних. Поняття «загальнодоступних комп'ютерних послуг» («computing as public utility»)

[1, с. 236], що належить Дж. Маккарті, з'явилось саме у збірнику «Управління та комп'ютери майбутнього» за редакцією М. Грінбергера. Подальший розвиток – від комп'ютерних послуг до інформаційних – став предметом обговорення в роботі «Комп'ютери завтрашнього дня» [7].

М. Грінбергер зазначає, що економічні і політичні умови, що склались у 1960-х роках, сприяють поширенню комп'ютерних послуг у всіх сферах суспільного життя для розв'язання широкого кола завдань: від рутинних чисельних розрахунків і маніпуляції текстовими даними до автоматичного управління приладами, моделювання динамічних процесів, статистичного аналізу та забезпечення інформаційної діяльності.

Необхідною основою цього є поширення систем розподілу комп'ютерного часу, що надають можливість зменшити накладні витрати, пов'язані з опрацюванням користувацьких програм і даних, їх зберіганням і доставлянням.

Процедура надання комп'ютерних послуг має бути простою і прозорою, як надання послуг електричної мережі:

– по-перше, щоб отримати доступ до електричної мережі, достатньо ввімкнути клему на комутаторі або вставити вилку в розетку – так само простим має бути доступ й до комп'ютерної мережі;

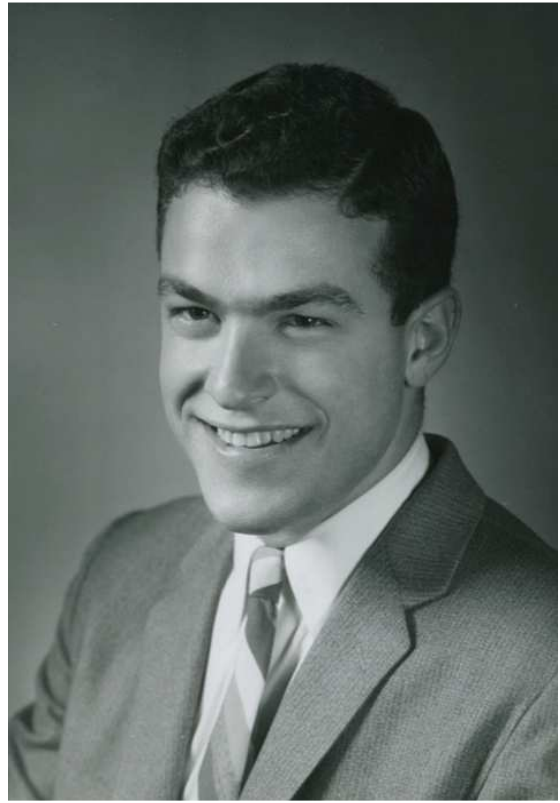

*Мартін Грінбергер як викладач Слоунівської школи менеджменту Массачусетського технологічного інституту (фотографія за 1959 рік з Музею МІТ)*

– по-друге, була винайдена і вдосконалена велика кількість різних видів електроустаткування, кожне з яких виконує власну функцію і має власне живлення – так само і комп'ютер з ЕОМ загального призначення має еволюціонувати у велику кількість різних видів комп'ютерних пристроїв, пов'язаних комп'ютерною мережею;

– по-третє, електрика є відносно однорідним продуктом, що централізовано виробляється і безперервно надається споживачеві – так само й комп'ютерні послуги мають бути стандартизовані та надаватись споживачеві за вимогою.

Головним різновидом комп'ютерних послуг М. Грінбергер вважає інформаційні послуги загального призначення, особливістю яких є те, що їх виробництво і споживання відбувається в одній мережі і виконується одними засобами, на відміну від доступу до електричної мережі й виробництва електроенергії.

Ураховуючи активний розвиток нинішніх ліній зв'язку, М. Грінбергер вважав малойомовірним, що компанії, які надаватимуть інформаційні послуги, будуть інвестувати у створення власних мереж зв'язку. Цей прогноз здійснювався протягом майже 50 років, доки провідні світові постачальники інформаційних послуг не розпочали розвиток власних безпровідних мереж зв'язку на основі аеростатів (Google [10]) і безпілотників (Facebook [14]) з метою забезпечення доступу до послуг мешканцям сільських, віддалених та необслуговуваних районів.

Забезпечення надання інформаційних послуг може бути державним (сучасними прикладами такого забезпечення є державні Інтернет-провайдери КНДР), державно регульованим (приклад – фільтрація Інтернет-трафіку з метою забезпечення інформаційної безпеки КНР) та вільним (нерегульованим).

Від інформаційних послуг важко або взагалі неможливо відмовитись, проте для їх надання необхідні великі інвестиції у програмне забезпечення послуг. Частину цих витрат можна перекласти на клієнтів, але тоді відповідальність за розробку, обслуговування та модифікацію ядра системи надання послуг необхідно покласти на постачальника послуги.

Найпершою сферою упровадження інформаційних послуг, на думку М. Грінбергера, будуть банківська справа і роздрібна торгівля. Автор фактично описує сучасний стан мобільного й Інтернет-банкінгу, роботу касових терміналів тощо, наголошуючи на тих зручностях, що вони надають. Прогноз щодо поширення віртуальних магазинів та електронних грошей не просто справдився – нині це є провідним напрямом розвитку торгівельно-грошових відносин.

Серед інших прикладів, що обговорюються в роботі – комп'ютерно керовані страхові послуги і фондові ринки («комп'ютер може бути запрограмований для підтримки стабільного та динамічного ринку ... в інтересах суспільства. Кожен інвестор матиме "місце" на комп'ютеризованому обміні, і навіть брокери стануть необов'язковими...» [7]), медико-інформаційні системи, системи централізованого контролю дорожнього трафіку, системи комп'ютерної торгівлі, автоматичні бібліотеки, інтегровані системи управління, навчальні термінали та навчальні мережі, науково-дослідні термінали в лабораторіях, проектні термінали в інженерних фірмах, комп'ютеризовані спільноти тощо.

Окремий розділ роботи [7] присвячено наданню інформаційних послуг з моделювання як найбільш перспективного підходу до аналізу складних систем і випадкових процесів: «використання моделювання ... системними аналітиками, керівниками, вченими-суспільствознавцями, та іншими фахівцями буде помітно розширюватися, оскільки інформаційна послуга забезпечує легкий доступ до потужних комп'ютерів і систем програмування.

Більшість користувачів [інформаційних послуг з] моделювання не матимуть знання або бажання будувати свої власні моделі... Сприяння в розробці, налагодженні, та перевірці достовірності моделей буде надана он-лайн центром моделювання...

Якщо виключити непередбачені перешкоди, інтерактивний он-лайн комп'ютерний сервіс з надання інформаційних послуг може бути настільки ж звичайним явищем у 2000 році, як телефонний зв'язок сьогодні. У 2000 році людина повинна мати набагато краще розуміння себе і своєї системи, не тому, що вона буде розумнішою за наших сучасників, а тому, що вона навчиться творчо використовувати найпотужніший з винайдених підсилювачів інтелекту.» [7]

Через 35 років ідеї А. О. Манна щодо надання комп'ютерних послуг як комунальних реалізувались Інтернет-провайдерами, а ще через 15 років ідеї М. Грінбергера стосовно надання спеціалізованих комп'ютерних інформаційних послуг (зокрема, з комп'ютерного моделювання) через мережі зв'язку трансформувались у хмарні технології, основою яких є хмарні обчислення (cloud computing).

У 2014 році А. В. Звєрєва поняття послуги системної інтеграції у сфері інформаційних технологій розглядає як «сукупність ІТ-послуг по створенню та супроводу (післяпродажному обслуговуванню) ІТ-інфраструктури підприємства, яка охоплює центри опрацювання даних, мережі, робочі місця, прикладне програмне забезпечення та корпоративні інформаційні системи, а також систему інформаційної безпеки» й класифікує цю послугу за такими основними ознаками:

– за територіальною ознакою: послуги локальні (обмежені підприємством) або територіально розподілені;

– за масштабом виконуваного проекту: створення комплексного інформаційного середовища або автоматизація окремих видів діяльності організації в рамках уже наявного інформаційного середовища;

– за наявності / відсутності аутсорсингу: побудова інформаційного простору на базі власної ІТ-інфраструктури або різні види оренди ІТ-інфраструктури і супутніх послуг;

– за місцем надання послуги: послуга призначена для об'єктів, розташованих у великих містах, регіональних центрах, невеликих містах і малих населених пунктах [19, с. 12–14].

За цими ознаками послуга системної інтеграції у сфері інформаційних технологій є спеціалізованою комп'ютерною інформаційною послугою, що надається засобами хмарних обчислень, перше повідомлення про які, за даними Google Books Ngram Viewer [6], з'явилось у грудні 2007 року (рис. 1).

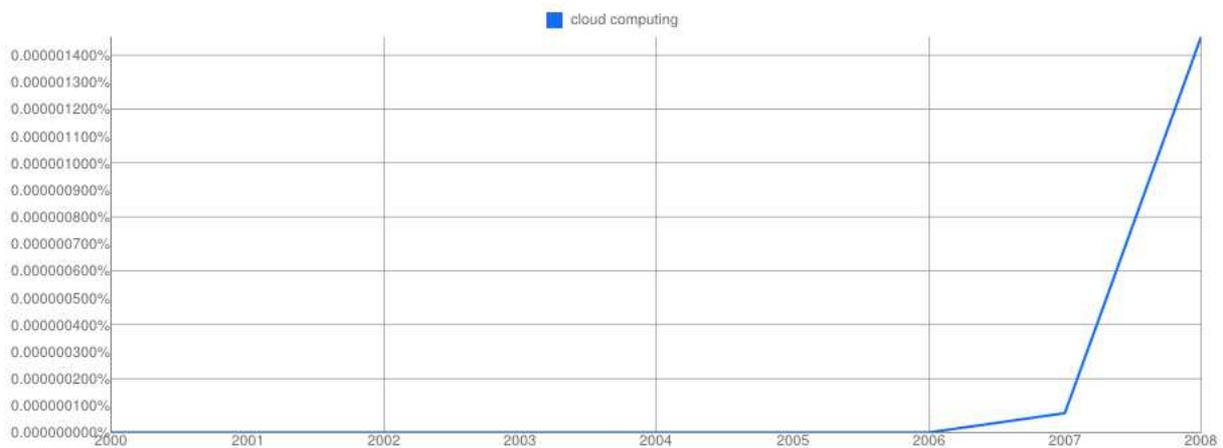

*Рис. 1. Результати пошуку в системі Google Books Ngram Viewer*

Авторами першого повідомлення про хмарні обчислення «Google and IBM Donate 1,600 Computers to 'Cloud' Project» (рис. 2) наголошувалось про існуючу можливість узгодженого й ефективного розв'язання спільної проблеми великою кількістю комп'ютерів, у зв'язку з чим низка компаній та університетів США виділили кошти на створення і застосування «хмарних» обчислювальних проектів: «За суттю, хмари є кластерами комп'ютерів – чисельністю від декількох десятків до декількох тисяч – що опрацьовують дані одночасно. Google та IBM щойно пожертвували 1600 комп'ютерів для використання з цією метою у трьох університетах. Один з них, Університет штату Меріленд – планує використовувати хмару для перекладу складних іншомовних текстів. Студенти створять програмне забезпечення для переваг хмарного комп'ютера. Учасники вірять, що така підготовка необхідна для того, щоб упоратись зі зростаючим обсягом даних, що мають бути опрацьовані.» [5]

У дисертації Д. Е. Ірвіна (David E. Irwin), присвяченій архітектурі операційних систем для інфраструктури мережного серверу, поняття «cloud computing» ототожнюється з «utility computing» [9, с. 38].

Р. М. Шор (R. M. Shor) вказує, що організація хмарних обчислень потребує доступу до Інтернету, серверів, високошвидкісних мереж, дискового простору, баз даних та програмного забезпечення загального призначення [15, с. 3]. На думку Д. Сігле (Del Siegle), використання хмарних обчислень надає студентам і викладачам основні переваги розглянутих раніше комп'ютерних (інформаційних) послуг, зокрема – можливість використовувати програми без встановлення їх на свої комп'ютери, а також забезпечує доступ до збережених файлів з будь-якого комп'ютера, підключеного до Інтернету. Як сучасна «комп'ютерна утиліта», хмарні технології забезпечують: більш ефективні обчислення за рахунок централізованого зберігання, опрацювання та високої пропускної здатності, одночасну роботу над проектом великої кількості користувачів незалежно від їх місцезнаходження. Термін «хмарні обчислення» використовується тому, що послуги і зберігання надаються через Інтернет (або хмару) [16].

> **Google and IBM Donate 1,600 Computers to 'Cloud' Project**
>
> It stands to reason you would be able to tackle a task more effectively with many computers than you could with one—the challenge is getting the many to work in concert. That's why a number of U.S companies and universities are devoting resources to the creation and application of "cloud" computing projects. Clouds are essentially clusters of computers—numbering from dozens to thousands—that process data simultaneously.
>
> Google and IBM recently donated 1,600 computers to be used for this purpose to three universities. One, the University of Maryland, plans to use its cloud to translate difficult foreign language texts. Students there will write the software to take advantage of the cloud computer. Participants believe such training is essential for keeping pace with the growing amount of data needing to be processed.

*Рис. 2. Перше повідомлення про хмарні обчислення [5]*

Незважаючи на наступність концепцій інформаційної послуги і хмарних обчислень, забезпечення останніх стимулювало розвиток низки нових професій, таких як архітектор бізнес та інформаційних технологій, фахівець з хмарної міграції, інженер-програміст хмаро орієнтованого програмного забезпечення, експерт з безпеки даних, мережний та комунікаційний аналітик [15, с. 16].

П. Мелл (Peter Mell) та Т. Гранц (Timothy Grance) визначають хмарні обчислення як модель надання, за необхідності, повсюдного і зручного мережного доступу до спільно використовуваних налаштовуваних обчислювальних ресурсів, які можуть бути швидко надані й вивільнені з мінімальними зусиллями з управління або з взаємодії з постачальником послуг (сервіс-провайдером) [13, с. 2].

Таке трактування хмарних обчислень більш відоме як «визначення NIST» (The NIST Definition of Cloud Computing). Згідно цього визначення, хмарна модель підтримує високу доступність послуг (сервісів), описується п'ятьма основними характеристиками (самообслуговування на вимогу, широкий мережний доступ, об'єднання ресурсів у пули, швидка еластичність та вимірюваність послуги), регулюється чотирма моделями розгортання (приватна хмара, хмара співтовариства, публічна хмара та гібридна хмара).

Автори [13] пропонують три моделі надання хмарних послуг:
1) програмне забезпечення як послуга – Cloud Software as a Service (SaaS);
2) платформа як послуга – Cloud Platform as a Service (PaaS);

3) інфраструктура як послуга – Cloud Infrastructure as a Service (IaaS).

Сучасним розвитком моделі SaaS є DaaS (Desktop as Service), за якої постачальник послуг надає споживачеві доступ до віртуального екрану програмного засобу (робочого стола тощо), що виконуються в хмарній інфраструктурі.

Визначення і моделі NIST значною мірою базуються на матеріалах «Білої книги хмарних технологій» [11] Syntec informatique (Франція), у якій вперше виконано порівняння класичної і хмарних моделей надання комп'ютерних послуг (рис. 3). Класична модель надання комп'ютерних послуг відзначається тим, що споживач послуги повністю управляє інфраструктурою, у той час як хмарні моделі розташовані на рис. 3 у порядку зменшення керованості з боку споживача.

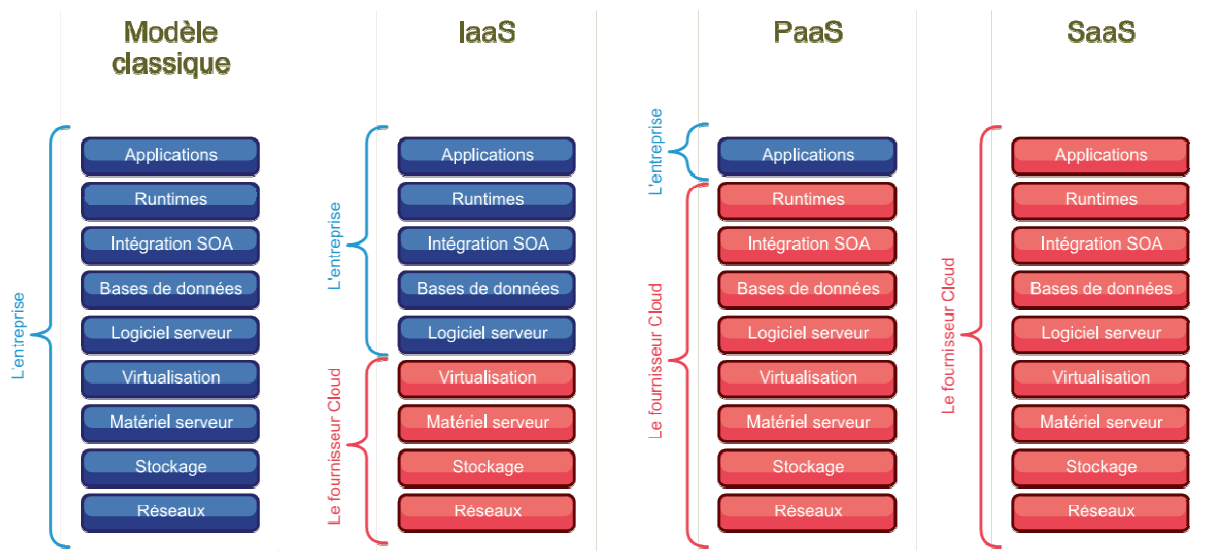

*Рис. 3. Класичні та хмарні моделі надання комп'ютерних послуг [11, с. 6]*

Як стверджує Д. Сігле, «економічно доведено, що хмарні обчислення економлять час і гроші. Як недорогий та освітньо корисний ресурс, ці технології будуть корисні як для школярів, так і для студентів, оскільки надають можливість пізнання різних комп'ютерних досягнень та здобуття навичок роботи у спільному навчальному середовищі. Так роботу, розпочату над проектом у навчальному закладі, можна продовжити вдома, якщо здійснити передачу файлів, завантаживши спільне програмне забезпечення» [16, с. 42].

Для визначення хмарних технологій навчання скористаємося еталонним трактуванням М. І. Жалдака *інформаційно-комунікаційних технологій* як сукупності «методів, засобів і прийомів, використовуваних для збирання, систематизації, зберігання, опрацювання, передавання, подання всеможливих повідомлень і даних» [18, с. 8].

Тоді *хмарні технології* (хмарні інформаційно-комунікаційні технології) як різновид ІКТ можна визначити як сукупність методів, засобів і прийомів, використовуваних для збирання, систематизації, зберігання та опрацювання на віддалених серверах, передавання через мережу і подання через клієнтську програму всеможливих повідомлень і даних.

За В. Ю. Биковим, педагогічна технологія – це структура організації часової і просторової взаємодії складових педагогічної системи і компонент навчальної системи, яка побудована відповідно до цілей навчання і виховання, змісту навчання та обраних методів навчання і виховання [17, с. 386]. Множину педагогічних технології, що використовуються у процесі навчання, називають *технологіями навчання*.

За В. Ю. Биковим, *інформаційно-комунікаційні технології навчання* – це «комп'ютерно орієнтована складова педагогічної технології, яка відображає деяку формалізовану модель певного компоненту змісту навчання і методики його подання у навчальному процесі, яка представлена в цьому процесі педагогічними програмними засобами і яка передбачає використання комп'ютера, комп'ютерно орієнтованих засобів навчання і комп'ютерних комунікаційних мереж для розв'язування дидактичних завдань або їх фрагментів» [17, с. 141].

Ураховуючи, що хмарні технології є підмножиною інформаційно-комунікаційних технологій, а ІКТ навчання є підмножиною технологій навчання, під *хмарними технологіями навчання* будемо розуміти такі ІКТ навчання, що передбачають використання хмарних ІКТ. Останні спрощено можуть бути визначені як мережні ІКТ, що передбачають централізоване мережне зберігання й опрацювання даних (виконання програм), за якого користувач виступає клієнтом (користувачем послуг), а «хмара» – сервером (постачальником послуг).

На рис. 4 наведено співвідношення змісту понять «технологія навчання», «ІКТ», «хмарні технології», «ІКТ навчання» та «хмарні технології навчання».

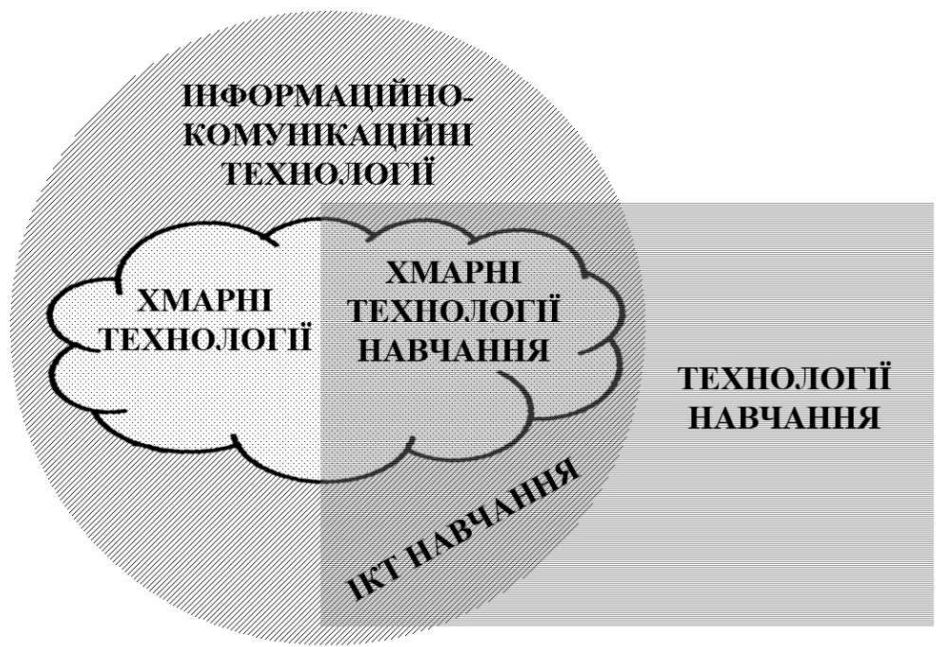

*Рис. 4. Співвідношення технологій навчання, інформаційно-комунікаційних технологій, хмарних технологій, ІКТ навчання та хмарних технологій навчання (▨ – технології навчання, ▨ – ІКТ, ▨ – ІКТ навчання, ▨ – хмарні технології, ▨ – хмарні технології навчання)*

## 3. ВИСНОВКИ ТА ПЕРСПЕКТИВИ ПОДАЛЬШИХ ДОСЛІДЖЕНЬ

1. Хмарні технології (хмарні ІКТ) як різновид ІКТ можна визначити як сукупність методів, засобів і прийомів, використовуваних для збирання, систематизації, зберігання та опрацювання на віддалених серверах, передавання через мережу і подання через клієнтську програму всеможливих повідомлень і даних. Витоки хмарних технологій навчання містяться у застосуванні концепції комп'ютерної послуги до навчального процесу, зокрема, надання місця для зберігання електронних освітніх ресурсів та мобільного доступу до них.

2. Перше систематичне дослідження проблеми надання комп'ютерних послуг було виконане у 1959 році А. О. Манном, який визначив, що найбільш перспективною комп'ютерною послугою є доступ до комунальних мереж передавання й опрацювання даних на основі великих, потужних, швидкісних центрів опрацювання даних. Зважаючи на появу засобів персональних обчислень, зростання характеристик яких може призвести до відсутності необхідності в комунальних обчисленнях (як це сталось з появою персональних комп'ютерів), А. О. Манн наголосив, що саме зростаюча складність зв'язків і комунікацій змусить об'єднуватись і їх користувачів для ефективного опрацювання даних.

3. За А. О. Манном, надання комп'ютерних послуг створює нові можливості для організацій (через можливість споживання більшої кількості комп'ютерних послуг меншої вартості), постачальників комп'ютерного обладнання (через перспективи для збільшення швидкості, ємності, потужності та переходу до безпаперового документообігу), розробників програмних систем (через засоби уніфікації використовуваних мов програмування, трансляторів тощо) і користувачів (через рівний доступ до програм і даних, розв'язання проблем немобільності програм та обладнання, застосування сучасних технологій опрацювання даних та інших засобів задоволення їх зростаючих потреб).

4. Уже через п'ять років технологічні проблеми надання комп'ютерних послуг, указані А. О. Манном, відійшли на другий план перед соціальними, визначеними Д. Ф. Паркхіллом і М. Грінбергером: загрози конфіденційності, технологічне безробіття, можливості для політичного контролю і багатьма іншими небезпеками, потенційними у 1960-х р. і реальними у 2010-х. Саме їх подолання фахівці Gartner визначають як перспективні напрями розвитку хмарних технологій.

5. Значний внесок у систематизацію термінології і моделей надання хмарних послуг зроблено європейськими фахівцями, зокрема – Syntec informatique (Франція), у той час як визначення й моделі, наведені у рекомендаціях NIST, є переважно стереотипними до більш ранніх досліджень.

6. Розвиток концепції комп'ютерних послуг за останні 50 років відзначався неперервністю: термінальні послуги 1960-х викликали до життя мережні операційні системи 1970-х, у яких сформувалась мережна культура 1980-х, що набула поширення у гіпертекстовому Інтернет 1990-х, який перетворився на джерело надання послуг у 2000-х та компонент соціального життя – у 2010-х. У той же час розвиток засобів хмарних технологій є діалектичним: у 2010-х, як і в 1980-х рр., нові термінальні комп'ютерні пристрої (смартфони, фаблети тощо) все більше набувають характеристик персональних комп'ютерів із мережним доступом, що можуть об'єднуватись у власні мережі за p2p-технологіями (Bluetooth, WiFi Direct тощо).

7. Виокремлення усталених і перспективних хмарних технологій в освіті вимагає наукового прогнозування розвитку хмарних технологій навчання на основі урахування історичних тенденцій розвитку ІКТ.

## СПИСОК ВИКОРИСТАНИХ ДЖЕРЕЛ

1. Computers and the World of the Future / Edited by Martin Greenberger. – New York : M.I.T. Press and Wiley, 1962. – 340 p.
2. Fenn J. Gartner's Hype Cycle Special Report for 2009 [Electronic resource] / Jackie Fenn, Mark Raskino, Brian Gammage // Gartner, Inc. and/or its Affiliates. – 2009. – Mode of access : http://www.gartner.com/id=1108412.

3. Flood M. M. DOUGLAS F. PARKHILL, The Challenge of the Computer Utility, Addison-Wesley Publishing Co., Reading, Mass., 1966, 219 pages, $7.95 / Merrill M. Flood // Operations Research. – 1967. – Vol. 15. – Issue 1. – P. 177–178.
4. Garfinkel S. L. Architects of the Information Society: 35 Years of the Laboratory for Computer Science at MIT / Simson L. Garfinkel ; edited by Hal Abelson. – Cambridge : The MIT Press, 1999. – 72 p.
5. Google and IBM Donate 1,600 Computers to 'Cloud' Project [Electronic resource] // Maximum PC. – 2007. – December. – P. 10. – Access mode : http://dl.maximumpc.com/Archives/MPC1207-web.pdf
6. Google Books Ngram Viewer [Electronic resource] / Google. – 2013. – Access mode : https://books.google.com/ngrams.
7. Greenberger M. The Computers of Tomorrow [Electronic resource] / Martin Greenberger // The Atlantic Monthly. – 1964. – Vol. 213. – No 5, May. – P. 63–67. – Access mode : http://www.theatlantic.com/past/docs/unbound/flashbks/computer/greenbf.htm.
8. Haigh T. Software in the 1960s as Concept, Service, and Product / Thomas Haigh // IEEE Annals of the History of Computing. – 2002. – Vol. 24. – Issue No. 1, January–March. – P. 5–13.
9. Irwin D. E. An Operating System Architecture for Networked Server Infrastructure : Dissertation submitted in partial fulfillment of the requirements for the degree of Doctor of Philosophy in the Department of Computer Science in the Graduate School of Duke University / David E. Irwin ; Department of Computer Science, Duke University. – [December], 2007. – XVII, 193 p.
10. Lardinois F. Google X Announces Project Loon: Balloon-Powered Internet For Rural, Remote And Underserved Areas [Electronic resource] / Frederic Lardinois // TechCrunch / AOL Inc. – Jun 14, 2013. – Access mode : http://techcrunch.com/2013/06/14/google-x-announces-project-loon-balloon-powered-internet-for-rural-remote-and-underserved-areas/.
11. Le Livre Blanc du Cloud Computing: Tout ce que vous devez savoir sur l'informatique dans le nuage / Syntec informatique. – 2$^{ème}$ Trimestre 2010. – 19 s.
12. Mann A. O. A publicly regulated system of management control services / Alan O. Mann // Management control systems : the proceedings of a symposium held at System Development Corporation, Santa Monica, California, July 29-31, 1959 / Edited by : Donald G. Malcolm, and Alan J. Rowe ; general editor : Lorimer F. McConnell. – Third printing. – New York ; London : John Wiley & Sons, 1962. – P. 245-263.
13. Mell P. The NIST Definition of Cloud Computing : Recommendation of the National Institute of Standards and Technology [Electronic resource] / Peter Mell, Timothy Grance. – Gaitherburg : National Institute of Standards and Technology, September 2011. – III, 3 p. – (Special Publication 800-415). – Access mode : http://csrc.nist.gov/publications/nistpubs/800-145/SP800-145.pdf.
14. Perez S. Facebook Looking Into Buying Drone Maker Titan Aerospace [Electronic resource] / Sarah Perez, Josh Constine // TechCrunch / AOL Inc. – Mar 3, 2014. – Access mode : http://techcrunch.com/2014/03/03/facebook-in-talks-to-acquire-drone-maker-titan-aerospace/.
15. Shor R. M. Cloud Computing for Learning and Performance Professionals / R. M. Shor // INFOLINE. – 2011. – April. – Issue 1104. – 22 p.
16. Siegle D. Cloud Computing: A Free Technology option to Promote Collaborative learning / Del Siegle // Gifted Child Today. – 2010. – Fall. – Vol. 33, No 4. – P. 41–45.
17. Биков В. Ю. Моделі організаційних систем відкритої освіти : [монографія] / В. Ю. Биков. – К. : Атіка, 2009. – 684 с. : іл.
18. Жалдак М. І. Проблеми інформатизації навчального процесу в середніх і вищих навчальних закладах / М. І. Жалдак // Комп'ютер в школі та сім'ї. – 2013. – № 3. – С. 8–15.
19. Зверева А. В. Формирование маркетинга услуг системной интеграции на основе облачных технологий : автореф. дисс. ... канд. эконом. наук : 08.00.05 «Экономика и управление народным хозяйством: маркетинг» / Зверева Анна Владимировна ; [ФГОБУВПО «Финансовый университет при Правительстве Российской Федерации»]. – М., 2014. – 29 с.
20. Облачные технологии и образовании / Сейдаметова З. С., Аблялимова Э. И., Меджитова Л. М., Сейтвелиева С. Н., Темненко В. А. : под общ. ред. проф. З. С. Сейдаметовой. – Симферополь : ДИАЙПИ, 2012. – 204 с.
21. Хмарні обчислення [Електронний ресурс] / Lynxrv // Вікіпедія – вільна енциклопедія. – 22:59, 26 березня 2012. – Режим доступу : https://uk.wikipedia.org/w/index.php?title=Хмарні_обчислення&oldid=9137614#.D0.86.D1.81.D1.82.D0.BE.D1.80.D1.96.D1.8F.
22. Чорна О. В. Використання циклу надочікувань для виявлення тенденцій розвитку хмарних технологій / О. В. Чорна // Хмарні технології в освіті : матеріали Всеукраїнського науково-методичного Інтернет-семінару (Кривий Ріг – Київ – Черкаси – Харків, 21 грудня 2012 р.). – Кривий Ріг : Видавничий відділ КМІ, 2012. – С. 3–6.

# ОБЛАЧНЫЕ ТЕХНОЛОГИИ ОБУЧЕНИЯ: ПРОИСХОЖДЕНИЕ

**Маркова Оксана Николаевна**
старший преподаватель кафедры компьютерных систем и сетей
ГВУЗ «Криворожский национальный университет», г. Кривой Рог, Украина
*kissa_oks@mail.ru*

**Семериков Сергей Алексеевич**
профессор, доктор педагогических наук,
заведующий кафедрой фундаментальных и социально-гуманитарных дисциплин
ГВУЗ «Криворожский национальный университет», г. Кривой Рог, Украина
*semerikov@gmail.com*

**Стрюк Андрей Николаевич**
кандидат педагогических наук, докторант
Институт информационных технологий и средств обучения НАПН Украины, г. Киев, Украина
*andrey.n.stryuk@gmail.com*

**Аннотация.** Целью исследования является рассмотрение эволюции концепции компьютерной услуги в работах зарубежных исследователей 1959–66 гг. Впервые введены в отечественный научный оборот результаты А. О. Манна и расширен обзор результатов Д. Ф. Паркхилла по концепции компьютерной (информационной) услуги. Доказана функциональная идентичность концепции компьютерной и облачной услуги, уточнено первоисточники моделей предоставления облачных услуг. Приведена трактовка понятия «облачные технологии обучения». Сделан вывод о непрерывности развития облачных технологий за последние 55 лет и их связь с развитием ИКТ в целом. Результаты исследования дают возможность определить перспективы развития облачных технологий в целом и облачных технологий обучения в частности.

**Ключевые слова:** компьютерная услуга; коммунальные вычисления; облачные технологии; облачные технологии обучения.

# THE CLOUD TECHNOLOGIES OF LEARNING: ORIGIN

**Oksana M. Markova**
senior lecturer of the Department of Computer Systems and Networks
SIHE "Kryvyi Rih National University", Kryvyi Rih, Ukraine
*kissa_oks@mail.ru*

**Serhiy O. Semerikov**
Full Professor, D. Sc. (pedagogical sciences),
head of the Department of Fundamental and Sociohumanitarian Disciplines
SIHE "Kryvyi Rih National University", Kryvyi Rih, Ukraine
*semerikov@gmail.com*

**Andrii M. Striuk**
Ph. D. (pedagogical sciences), doctorate student
Institute of Information Technologies and Learning Tools, Kyiv, Ukraine
*andrey.n.stryuk@gmail.com*

**Abstract.** The research goal is to investigate the evolution of the concept of utility computing in the works of foreign researchers in the years 1959-1966. First the A. O. Mann's results and expanded overview of the D. F. Parkhill's results on the concept of computer (information) utility were introduced in the domestic scientific circulation. Functionally identity of the computer utility and cloud computing concepts was proved, as well as refined the primary sources of cloud service models. There was proposed the interpretation of the "cloud technologies of learning" concept. Continuity of the development of cloud technologies over the past 55 years and their relationship with the development of ICT in general was concluded. The research results make it possible to determine the prospects of the development of cloud computing in general and cloud technologies of learning in particular.

**Keywords:** computer utility; utility computing; cloud computing; cloud technologies of learning.

# REFERENCES (TRANSLATED AND TRANSLITERATED)

1. Computers and the World of the Future / Edited by Martin Greenberger. – New York : M.I.T. Press and Wiley, 1962. – 340 p. (in English).
2. Fenn J. Gartner's Hype Cycle Special Report for 2009 [online] / Jackie Fenn, Mark Raskino, Brian Gammage // Gartner, Inc. and/or its Affiliates. – 2009. – Available from : http://www.gartner.com/id=1108412 (in English).
3. Flood M. M. Douglas F. Parkhill, The Challenge of the Computer Utility, Addison-Wesley Publishing Co., Reading, Mass., 1966, 219 pages, $7.95 / Merrill M. Flood // Operations Research. – 1967. – Vol. 15. – Issue 1. – P. 177–178 (in English).
4. Garfinkel S. L. Architects of the Information Society: 35 Years of the Laboratory for Computer Science at MIT / Simson L. Garfinkel ; edited by Hal Abelson. – Cambridge : The MIT Press, 1999. – 72 p. (in English).
5. Google and IBM Donate 1,600 Computers to 'Cloud' Project [online] // Maximum PC. – 2007. – December. – P. 10. – Available from : http://dl.maximumpc.com/Archives/MPC1207-web.pdf (in English).
6. Google Books Ngram Viewer [online] / Google. – 2013. – Available from : https://books.google.com/ngrams (in English).
7. Greenberger M. The Computers of Tomorrow [online] / Martin Greenberger // The Atlantic Monthly. – 1964. – Vol. 213. – No 5, May. – P. 63–67. – Available from : http://www.theatlantic.com/past/docs/unbound/flashbks/computer/greenbf.htm (in English).
8. Haigh T. Software in the 1960s as Concept, Service, and Product / Thomas Haigh // IEEE Annals of the History of Computing. – 2002. – Vol. 24. – Issue No. 1, January–March. – P. 5-13. (in English)
9. Irwin D. E. An Operating System Architecture for Networked Server Infrastructure : Dissertation submitted in partial fulfillment of the requirements for the degree of Doctor of Philosophy in the Department of Computer Science in the Graduate School of Duke University / David E. Irwin ; Department of Computer Science, Duke University. – [December], 2007.— XVII, 193 p. (in English).
10. Lardinois F. Google X Announces Project Loon: Balloon-Powered Internet For Rural, Remote And Underserved Areas [online] / Frederic Lardinois // TechCrunch / AOL Inc. – Jun 14, 2013. – Available from: http://techcrunch.com/2013/06/14/google-x-announces-project-loon-balloon-powered-internet-for-rural-remote-and-underserved-areas/ (in English).
11. Le Livre Blanc du Cloud Computing: Tout ce que vous devez savoir sur l'informatique dans le nuage / Syntec informatique. –$2^{ème}$ Trimestre 2010. – 19 s. (in French) .
12. Mann A. O. A publicly regulated system of management control services / Alan O. Mann // Management control systems : the proceedings of a symposium held at System Development Corporation, Santa Monica, California, July 29-31, 1959 / Edited by : Donald G. Malcolm, and Alan J. Rowe ; general editor : Lorimer F. McConnell. – Third printing. – New York ; London : John Wiley & Sons, 1962. – P. 245–263 (in English).
13. Mell P. The NIST Definition of Cloud Computing : Recommendation of the National Institute of Standards and Technology [online] / Peter Mell, Timothy Grance. – Gaitherburg : National Institute of Standards and Technology, September 2011. – III, 3 p. – (Special Publication 800-415). – Available from : http://csrc.nist.gov/publications/nistpubs/800-145/SP800-145.pdf (in English).
14. Perez S. Facebook Looking Into Buying Drone Maker Titan Aerospace [online] / Sarah Perez, Josh Constine // TechCrunch / AOL Inc. – Mar 3, 2014. – Available from : http://techcrunch.com/2014/03/03/facebook-in-talks-to-acquire-drone-maker-titan-aerospace/ (in English).
15. Shor R. M. Cloud Computing for Learning and Performance Professionals / R. M. Shor // INFOLINE. – 2011. – April. – Issue 1104. – 22 p. (in English).
16. Siegle D. Cloud Computing: A Free Technology option to Promote Collaborative learning / Del Siegle // Gifted Child Today. – 2010. – Fall. – Vol. 33, No 4. – P. 41–45 (in English).
17. Bykov V. Yu. Models of the open education organizational systems : [monograph] / V. Yu. Bykov. – K. : Atika, 2009. – 684 p. : ill. (in Ukrainian).
18. Zhaldak M. I. Problems of informatization of educational process in secondary and higher education / M. I. Zhaldak // Kompyuter u shkoli ta simyi. – 2013. – Iss. 3. – P. 8–15. (in Ukrainian).
19. Zvereva A. V. Formation of the marketing system integration services based on cloud computing : Authoref. of Thesis ... Candidate in Econom. Sc. : 08.00.05 «Economics and National Economy

Management: Marketing» / Zvereva Anna Vladimirovna ; [FGONUVPO «Finansovyj universitet pri Pravitelstve Rossijskoj Federacii»]. – M., 2014. – 29 p. (in Russian).
20. Cloud technologies and education / Seydametova Z. S., Ablyalimova E. I., Medzhitova L. M., Seytvelieva S. N., Temnenko V. A. : ed. by prof. Z. S. Seydametova. – Simferopol : DIAYPI, 2012. – 204 p. (in Russian).
21. Cloud computing [online] / Lynxrv // Wikipedia – the free Encyclopedia. – 22:59, March 26, 2012. – Available from : https://uk.wikipedia.org/w/index.php?title=Хмарні_обчислення&oldid=9137614#.D0.86.D1.81.D1.82.D0.BE.D1.80.D1.96.D1.8F (in Ukrainian).
22. Tcshorna O. V. Using hype cycle to discover the trends of cloud technologies development / O. V. Tcshorna // Cloud technologies in education : proceedings of All-Ukrainian scientfic and methodic Internet-workshop (Kryvyi Rih – Kyiv – Cherkassy – Kharkiv, December 21, 2012). – Kryvyi Rih : Vydavnychyi viddil KMI, 2012. – P. 3–6 (in Ukrainian).